\documentclass[sigconf]{acmart}
\acmConference[MSR 2024]{MSR '24: Proceedings of the 21st International Conference on Mining Software Repositories}{April 15–16, 2024}{Lisbon, Portugal}
\AtBeginDocument{%
  }

\usepackage{algorithmic}
\usepackage{graphicx}
\usepackage{textcomp}
\usepackage{xcolor}

\usepackage{multirow}
\usepackage{booktabs}
\usepackage{graphics}

\copyrightyear{2024}
\acmYear{2024}
\setcopyright{acmlicensed}\acmConference[MSR '24]{21st International Conference on Mining Software Repositories}{April 15--16, 2024}{Lisbon, Portugal}
\acmBooktitle{21st International Conference on Mining Software Repositories (MSR '24), April 15--16, 2024, Lisbon, Portugal}
\acmDOI{10.1145/3643991.3648400}
\acmISBN{979-8-4007-0587-8/24/04}

\begin{document}

\title{DevGPT: Studying Developer-ChatGPT Conversations}

\author{Tao Xiao}
\orcid{0000-0003-4070-585X}
\affiliation{%
  \institution{Nara Institute of Science and Technology}
  \city{}
  \country{Japan}
}
\email{tao.xiao.ts2@is.naist.jp}

\author{Christoph Treude}
\orcid{0000-0002-6919-2149}
\affiliation{%
  \institution{University of Melbourne}
  \city{}
  \country{Australia}
}
\email{christoph.treude@unimelb.edu.au}

\author{Hideaki Hata}
\orcid{0000-0003-0708-5222}
\affiliation{%
  \institution{Shinshu University}
  \city{}
  \country{Japan}
}
\email{hata@shinshu-u.ac.jp}

\author{Kenichi Matsumoto}
\orcid{0000-0002-7418-9323}
\affiliation{%
  \institution{Nara Institute of Science and Technology}
  \city{}
  \country{Japan}
}
\email{matumoto@is.naist.jp}

\renewcommand{\shortauthors}{Xiao et al.}


\begin{abstract}
    This paper introduces DevGPT, a dataset curated to explore how software developers interact with ChatGPT, a prominent large language model (LLM). The dataset encompasses 29,778 prompts and responses from ChatGPT, including 19,106 code snippets, and is linked to corresponding software development artifacts such as source code, commits, issues, pull requests, discussions, and Hacker News threads. This comprehensive dataset is derived from shared ChatGPT conversations collected from GitHub and Hacker News, providing a rich resource for understanding the dynamics of developer interactions with ChatGPT, the nature of their inquiries, and the impact of these interactions on their work. DevGPT enables the study of developer queries, the effectiveness of ChatGPT in code generation and problem solving, and the broader implications of AI-assisted programming. By providing this dataset, the paper paves the way for novel research avenues in software engineering, particularly in understanding and improving the use of LLMs like ChatGPT by developers.
\end{abstract}



\maketitle

\section{High-level overview}
The emergence of large language models (LLMs) such as ChatGPT has disrupted the landscape of software development. Many studies are investigating the quality of responses generated by ChatGPT, the efficacy of various prompting techniques, and its comparative performance in programming contests, to name a few examples. Yet, we know very little about how ChatGPT is actually used by software developers. What questions do developers present to ChatGPT? What are the dynamics of these interactions? What is the backdrop against which these conversations are held, and how do the conversations feedback into the artifacts of their work? To close this gap, we introduce DevGPT, a curated dataset which encompasses 29,778 prompts and ChatGPT's responses including 19,106 code snippets, coupled with the corresponding software development artifacts---ranging from source code, commits, issues, pull requests, to discussions and Hacker News threads---to enable the analysis of the context and implications of these developer interactions with ChatGPT.


To create DevGPT, we leveraged a feature introduced by OpenAI in late May 2023, which allows users to share their interactions with ChatGPT through dedicated links.\footnote{\url{https://help.openai.com/en/articles/7925741-chatgpt-shared-links-faq}} We collected all such links shared on GitHub and Hacker News at nine specific points from July to October. If users chose to delete or deactivate their shared conversations in the intervening periods, we ensured data consistency by accessing the original shared link across all these snapshots.

\begin{table*}[t]
    \centering
    \caption{Summary Statistics of the snapshot 20231012}
    \label{tab:summary}
    \resizebox{\textwidth}{!}{
    \begin{tabular}{lrlrrrrr}
    \toprule
    \textbf{Sources} &  \textbf{\# } &  \textbf{Mentioned in} &  \multicolumn{3}{c}{\textbf{Shared ChatGPT Links}} & \multicolumn{2}{c}{\textbf{ChatGPT Conversations}} \\
    & & & \# Shared Links & \# Accessible Links & \# Conversations with Code & \# Prompts & \# Code Snippets 
    \\
    \midrule
    \multirow{1}{*}{\textbf{GitHub Code File}} & \multirow{1}{*}{1,843} & Code  & 2,708 & 2,540 & 1,184 & 22,799 & 14,132 \\
    \midrule
    \multirow{1}{*}{\textbf{GitHub Commit}} & \multirow{1}{*}{694} & Message  & 694 & 692 & 674 & 1,922 & 1,828 \\ 
    \midrule
    \multirow{3}{*}{\textbf{GitHub Issue}} & \multirow{3}{*}{507}
    &   Comment & 404 & 382 & 215 & 1,212 & 821\\
    & &  Description & 228 & 212 & 141 & 1,103 & 841 \\
    & & Title & 4 & 4 & 4 & 50 & 77 \\

    \midrule
    \multirow{3}{*}{\textbf{GitHub Pull Request}} & \multirow{3}{*}{267} & Description & 94 & 93 & 59 & 529 & 384\\
    & & Review Thread & 109 & 102 & 66 & 201 & 166 \\
    & & Comment & 98 & 91 & 54 & 430 & 425 \\
    \midrule
    \multirow{3}{*}{\textbf{Hacker News}} & \multirow{3}{*}{187} & Comment  & 267 & 234 & 44 & 849 & 127 \\
    & & Attached URL & 42 & 37 & 2 & 376 & 54 \\
    & & Story & 15 & 12 & 4 & 48 & 63 \\

    \midrule
    \multirow{3}{*}{\textbf{GitHub Discussion}} & \multirow{3}{*}{61} & Comment & 40 & 34 & 17 & 138 & 76 \\
    & & Description & 21 & 20 & 12 & 93 & 87 \\
    & & Reply & 9 & 7 & 5 & 28 & 25 \\

    \bottomrule
    \end{tabular}}

\end{table*}


Table~\ref{tab:summary} provides an overview of the snapshot 20231012. Comprising 4,733 shared ChatGPT links sourced from 3,559 GitHub or Hacker News references, the dataset contains a total of 29,778 prompts/answers. This includes 19,106 code snippets, with Python (6,084), JavaScript (4,802), and Bash (4,332) as the top three programming languages. 940 of these links are referenced across multiple sources, resulting in a unique count of 3,794 individual ChatGPT shared links within DevGPT.

\begin{figure*}[h]
\caption{Example of a ChatGPT conversation in the context of a GitHub pull request}
\label{fig:exm}
\centering
\includegraphics[width=\textwidth]{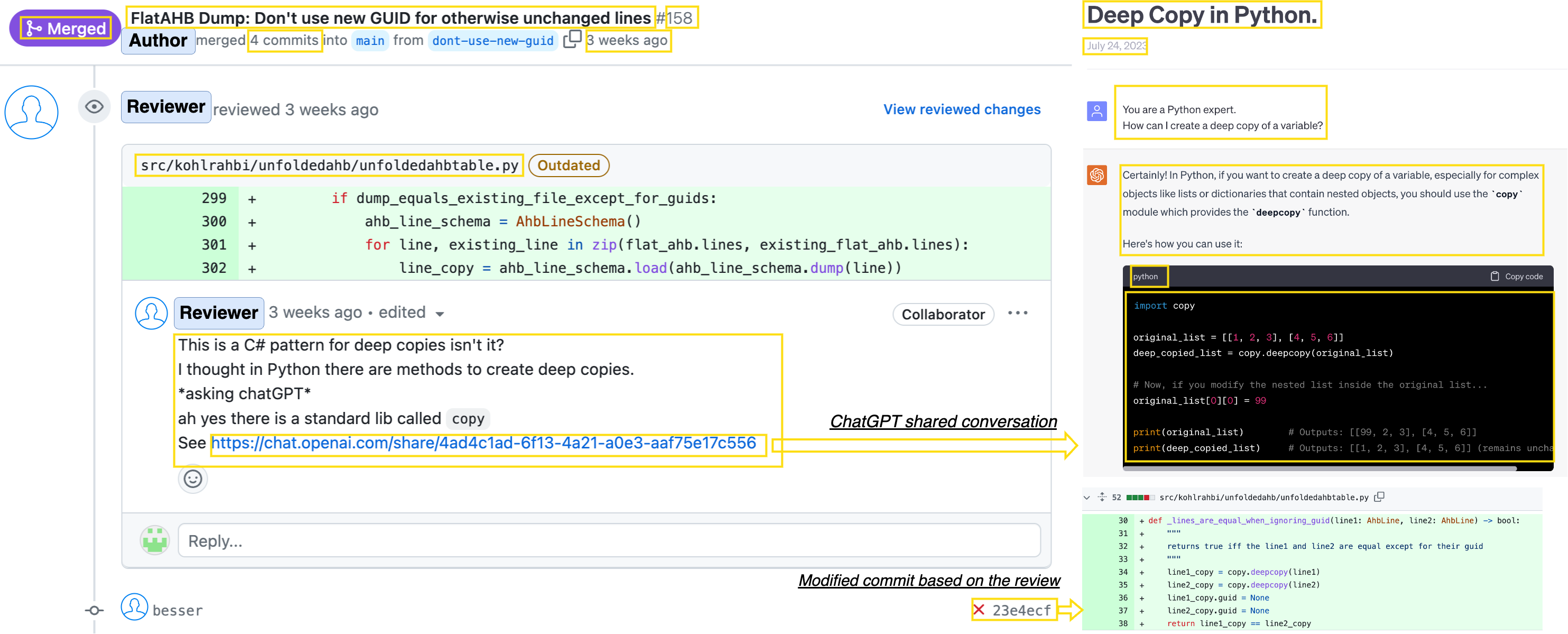}
\end{figure*}

Figure~\ref{fig:exm} shows an instance of a ChatGPT conversation from the dataset, together with the pull request it was related to and how the code was updated after the ChatGPT conversation.

\section{Internal structure}

The dataset consists of a collection of JSON files collected from the six sources detailed in Table~\ref{tab:summary}. For each source, we provide distinct metadata in the JSON file to enable source-specific analysis. Apart from the source-specific metadata, every JSON contains a consistent attribute: a list of shared ChatGPT links. Each shared link includes the URL to the ChatGPT conversation, the associated HTTP response status codes, the access date of the URL, and the content within the HTML response. Additionally, each conversation contains a list of prompts/answers, inclusive of any code snippets. We provide details including the date of the conversation, the count of prompts/answers, their token information, and the model version involved in the chat. Attributes detailing where the conversation was referenced are also included---such as the referencing URL, the nature of the mention (e.g., a comment), the individual who mentioned it, and the context in which it was cited. A comprehensive breakdown of the data structure is available at \url{https://github.com/NAIST-SE/DevGPT}. Additionally, we provide a CSV file cataloging all shared ChatGPT links gathered from GitHub and Hacker News.

\section{How to access}

The DevGPT dataset is available for download on Zenodo, see Section~\ref{sec:link}. It is formatted in JSON, making it easily parsable with any standard JSON library. Additionally, we include the HTTP response, which can be analyzed using any HTML parser. The dataset also categorizes code snippets by type, enabling researchers to use corresponding compilers for execution. No credentials are needed to access the dataset.

\section{Potential research questions}
The following provides a sample list of research questions that can be answered with the DevGPT dataset:
\begin{enumerate}
    \item What types of issues (bugs, feature requests, theoretical questions, etc.) do developers most commonly present to ChatGPT?
    \item Can we identify patterns in the prompts developers use when interacting with ChatGPT, and do these patterns correlate with the success of issue resolution?
    \item What is the typical structure of conversations between developers and ChatGPT? How many turns does it take on average to reach a conclusion?
    \item In instances where developers have incorporated the code provided by ChatGPT into their projects, to what extent do they modify this code prior to use, and what are the common types of modifications made?
    \item How does the code generated by ChatGPT for a given query compare to code that could be found for the same query on the internet (e.g., on Stack Overflow)?
    \item What types of quality issues (for example, as identified by linters) are common in the code generated by ChatGPT?
    \item How accurately can we predict the length of a conversation with ChatGPT based on the initial prompt and context provided?
    \item Can we reliably predict whether a developer's issue will be resolved based on the initial conversation with ChatGPT?
    \item If developers were to rerun their prompts with ChatGPT now and/or with different settings, would they obtain the same results?
\end{enumerate}

\section{Related work}
To situate the DevGPT dataset in the existing literature, in this section, we discuss existing research on link sharing and large language models (LLMs) in the field of software engineering.

\subsection{Link Sharing}
Link sharing, a prevalent method of knowledge sharing, is extensively adopted within developer communities, including Q\&A sites, GitHub, and code reviews.  G{\'o}mez et al. \cite{gomez2013study} found that a considerable number of links on Stack Overflow were used to share knowledge about software development innovations, such as libraries and tools. Ye et al.~\cite{ye2017structure} examined the structural and dynamic aspects of the knowledge network on Stack Overflow, noting that developers use links for various purposes, predominantly for referencing information to solve problems. Hata et al. \cite{hata20199} noted that over 80\% of repositories feature at least one link in source code comments. Xiao et al. \cite{xiao202318} expanded this research to include the role of links in commit messages, observing that inaccessible and patch links were most common. The practice of link sharing was also
studied in the context of code review. Zampetti et al. \cite{zampetti2017developers} explored the extent and purpose of external online resource references in pull requests, finding that developers often consult external resources to gain knowledge or resolve specific issues. Wang et al. \cite{wang2021understanding} employed a mixed-method approach to underscore the importance of shared links in review discussions, highlighting their role in satisfying the information needs of patch authors and review teams.

\subsection{LLMs for SE}
Since the introduction of the Transformer architecture in 2017 \cite{vaswani2017attention}, LLMs have become increasingly significant in Software Engineering (SE). Hou et al. \cite{hou2023large} conducted a systematic review of 229 research articles from 2017 to 2023, revealing the widespread use of LLMs in addressing software development problems. Prominent models in this area include \texttt{GPT-2/GPT-3/GPT-3.5} \cite{dong2023self, li2023enabling,liu2023improving,liu2023your,nascimento2023comparing,wang2023evaluating,yeticstiren2023evaluating}, \texttt{GPT-4} \cite{bareiss2022code,gilbert2023semantic,jiang2023selfevolve,liu2023your}, and the \texttt{BERT} series \cite{zeng2022extensive,lai2023ds}, demonstrating effectiveness in code generation, completion, and summarization.

Code completion, integral to Integrated Development Environments (IDEs) and code editors, has been enhanced by tools like \texttt{Codex} \cite{chen2021evaluating, doderlein2022piloting, li2023cctest, pearce2023examining}, the \texttt{BERT} series \cite{khan2022automatic}, \texttt{GitHub Copilot} \cite{doderlein2022piloting, li2023cctest, pudari2023copilot}, \texttt{CodeParrot} \cite{xu2022systematic,li2023cctest}, and the \texttt{GPT} series \cite{xu2022systematic,ochs2023evaluating}. Conversely, code summarization technologies like \texttt{Codex} \cite{ahmed2023improving, arakelyan2023exploring,gao2023constructing}, \texttt{CodeBERT} \cite{chen2022transferability,gu2022assemble,gao2023constructing}, and \texttt{T5} \cite{mastropaolo2021studying, mastropaolo2022using} focus on generating natural language descriptions from source code to facilitate maintenance, search, and classification.

In software maintenance, nearly a quarter of the studies reviewed by Hou et al. \cite{hou2023large} address program repair, code review, and debugging. In program repair, \texttt{Codex} \cite{wu2023effective,xia2023automated} and \texttt{ChatGPT} \cite{xia2023conversational} have shown strong performance. For code review, LLMs like \texttt{BERT} \cite{sghaier2023multi} and \texttt{ChatGPT} \cite{sridhara2023chatgpt} are effective in detecting issues and suggesting optimizations. Additionally, \texttt{Copilot for PRs} powered pull requests need less review time and have a higher likelihood of being merged~\cite{copilot}.

Despite these advances, there is limited research on how software developers interact with LLMs. The DevGPT dataset addresses this gap, offering a valuable resource for in-depth analysis of these interactions. This dataset can enable the research community to understand and improve the ways developers use LLMs in their work, marking a step forward in the practical application of AI in software development.

\section{Links}
\label{sec:link}
\url{https://github.com/NAIST-SE/DevGPT} and
\url{https://doi.org/10.5281/zenodo.10086809}
\bibliographystyle{ACM-Reference-Format}
\bibliography{main}

\balance

\end{document}